\pdfoutput=1  

\documentclass[aps,prd,superscriptaddress,floatfix,nofootinbib,preprintnumbers,
eqsecnum,
twocolumn]{revtex4-1}

\usepackage{amsmath,float,graphicx,placeins,amssymb,multirow,pgfplots,pgfplotstable}
\usepackage{tikz}
\usepackage{soul}
\setstcolor{red}

\usepackage{lipsum} 
\usepackage{comment}
\usepackage{enumerate,slashed,cancel,comment,color,bbm,graphicx,rotating,placeins,mathtools,multirow,hhline}
\usepackage[normalem]{ulem}
\usepackage{array}
\usepackage{enumitem}
\usepackage{hyperref}
\usepackage{color}
 \hypersetup
 { 
   colorlinks,
   linkcolor={blue!80!black},
   citecolor={blue!70!black},
   urlcolor={blue!70!black}
 }

\usepackage{diagbox}

\usepackage{subfig}
\usepackage{caption}

\usetikzlibrary{shapes.geometric,arrows}
\graphicspath{}

\def\ie{{\it i.e.}\/}

\begin{document}
\title{Enhancing the Energy Gap of Random Graph Problems via XX-catalysts in Quantum Annealing}

\def\andname{\hspace*{-0.5em}} 
\author{Luca A. Nutricati}
\email[Email address: ]{l.nutricati@ucl.ac.uk}
\affiliation{London Centre for Nanotechnology, University College London, WC1H 0AH London, UK}
\author{Roopayan Ghosh}
\email[Email address: ]{roopayan.ghosh@ucl.ac.uk}
\affiliation{Department of Physics and Astronomy, University College London, WC1E 6BT London, UK}
\author{Natasha Feinstein}
\email[Email address: ]{natasha.feinstein.19@ucl.ac.uk}
\affiliation{London Centre for Nanotechnology, University College London, WC1H 0AH London, UK}
\author{Sougato Bose}
\email[Email address: ]{s.bose@ucl.ac.uk}
\affiliation{Department of Physics and Astronomy, University College London, WC1E 6BT London, UK}
\author{P. A. Warburton}
\email[Email address: ]{p.warburton@ucl.ac.uk}
\affiliation{London Centre for Nanotechnology, University College London, WC1H 0AH London, UK}
\affiliation{Department of Electronic \& Electrical Engineering, University College London, WC1E 7JE London, UK}

\begin{abstract}
One of the bottlenecks in solving combinatorial optimisation problems using quantum annealers is the emergence of exponentially-closing energy gaps between the ground state and the first excited state during the annealing, which indicates that a first-order phase transition is taking place. The minimum energy gap scales inversely with the exponential of the system size, ultimately resulting in an exponentially large time required to ensure the adiabatic evolution. In this paper we demonstrate that employing multiple XX-catalysts on all the edges of a graph upon which a MWIS (Maximum Weighted Independent Set) problem is defined significantly enhances the minimum energy gap. Remarkably, our analysis shows that the more severe the first-order phase transition, the more effective the catalyst is in opening the gap. This result is based on a detailed statistical analysis performed on a large number of randomly generated MWIS problem instances on both Erdős–Rényi and Barab\a'asi-Albert graphs. We also observe that similar performance cannot be achieved by the non-stoquastic version of the same catalyst, with the stoquastic catalyst being the preferred choice in this context.

\end{abstract}
\maketitle

\tableofcontents

\def\beq{\begin{equation}}
\def\eeq{\end{equation}}

\section{Introduction}

Optimisation problems are of critical importance in numerous research fields, from protein folding in biology \cite{Levinthal}, to drug design in chemistry \cite{Aspuru,Lanyon,elfving2020quantum} and portfolio optimisation in finance \cite{fintech:1,fintech:2,fintech:3} as well as applications in high energy physics, see for example \cite{Allanach:2004my, Douglas:2006xy, Abel:2014xta, Halverson:2018cio, Cole:2019enn, Halverson:2019tkf, Ruehle:2020jrk, Larfors:2020ugo, Abel:2021rrj, Abel:2023zwg, Abel:2023rxo}. In particular, combinatorial optimisation problems in graph theory, such as maximum cut, maximum (weighted) independent set, maximum vertex cover and set cover (and generalisations thereof) have ubiquitous applications, from circuit layout design \cite{doi:10.1287/opre.36.3.493} to network routing, where data packets are arranged such that the structure can be modeled as an maximum weighted independent set problem \cite{ramaswami2009optical}. 

Given their application in many areas, it is crucial to develop techniques and possibly technologies whereby one can efficiently solve these problems, overcoming their intrinsic complexity from the classical perspective.  Many of them are NP-hard, meaning they cannot be classically solved in non-deterministic polynomial time. With the emergence of powerful near-term quantum devices and algorithms, it is natural to consider if these technologies could offer a qualitatively different solution for NP-hard problems. Specifically, could they provide a novel method to overcome classical techniques leading to a less harsh exponential scaling exponent? Although various quantum computing paradigms have been proposed, one type of quantum computer, known as a quantum annealer, has been designed specifically for the task of solving optimisation problems.

In current quantum annealers each qubit physically corresponds to a superconducting loop with an associated magnetic flux. Qubits are coupled together with various topologies using inductive couplers. Quantum annealers are machines able to evolve a quantum system initialising it using a local transverse field inducing all the qubits to be in a superposition of states, which corresponds to the ground state of the system. Gradually, using flux biases, the transverse field is shut down and the time-dependent Hamiltonian of the system evolves towards a new Hamiltonian whose ground-state encodes the optimal solution of the problem to be solved. If this evolution proceeds adiabatically, the ground-state of the final Hamiltonian is obtained with high probability at the end of the anneal.

Let us now emphasise some of the crucial aspects of quantum annealing which will be crucial in the following sections. The Hamiltonian that is implemented in available quantum annealers is usually written as
\begin{equation}
    H_A ~=~ (1-s)\,H_D \,+\, s \, H_P \, ,
\label{eq:annealing_hamiltonian}
\end{equation}
where $s = s(t)$ is called annealing parameter, which depends upon the physical time $t$ and defines the so-called {\it anneal schedule}. In the case of {\it forward annealing}, this is typically an arbitrary function which interpolates from $s=0$ at $t=0$ to $s=1$ at $t=T$, where $T$ is the total anneal time. This function has to be carefully chosen as in some cases its optimisation leads to dramatic performance improvements. However, this is beyond the scope of this paper and from now on  we shall focus on the simplest choice of this schedule, \ie, a linear function defined by
\begin{equation}
    s(t) ~=~ \frac{t}{T} \, .
\end{equation}
Finally, $H_D$ and $H_P$ denote the driver and the problem Hamiltonian, respectively. The driver Hamiltonian is a sum of local $X$-fields 
\begin{equation}
    H_D ~=~ -\sum_{i=1}^{N} \sigma_x^{(i)} \, ,
\label{eq:usual_driver}
\end{equation}
whereas the problem Hamiltonian encodes the problem we aim to solve in the ground of an Ising Hamiltonian
\begin{equation}
    H_P ~=~ \sum_{i,j \in E(G)} J_{ij} \, \sigma_z^{(i)} \sigma_z^{(j)} \,+\, \sum_{i \in V(G)} h_i \, \sigma_z^{(i)} \, ,
    \label{eq:problem}
\end{equation}
where $G$ is the particular graph that constitutes the Ising model, \ie, a set of $N$ nodes (or spins), indicated with $V(G)$, and links between them, indicated by $E(G)$ (couplings) with strength $J_{ij}$. Finally, $h_i$ are the local biases and $\sigma_z^{(i)}$ the Pauli $z$ matrix, which is related to the longitudinal spin state of the $i$-th spin.

As previously mentioned, quantum annealers make use of the adiabatic theorem to ensure that the entire evolution of the system from $s=0$ to $s=1$ is confined to the ground state of the annealing Hamiltonian in Eq.~\eqref{eq:annealing_hamiltonian}, without jumps to any excited state. This means that if the adiabatic condition is satisfied, the solution of our optimisation problem is encoded in the final classical values of the spins. However, the condition under which the adiabatic theorem holds dictates that the total anneal time must scale with the inverse of the minimum energy gap between the ground state and the first excited state, \ie,
\begin{equation}
    T ~\sim~ \frac{1}{\left[\min_{0 \leq s \leq 1}\Delta E_{01}(s)\right]^2} \,, 
    \label{eq:adiabatic_condition}
\end{equation}
where we recall that $T$ is the total anneal time and $\Delta E_{01}(s) = E_1(s) - E_0(s)$ is the energy gap between the ground and the first excited state of the annealing Hamiltonian $H_A$ for that specific value of the annealing parameter $s$. 

It is well known that some problems exhibit exponentially closing gaps in the system size, resulting in the following scaling with the total number of spins $N$:
\begin{equation}
    \min_{0 \leq s \leq 1}\Delta E_{01}(s) ~\sim~ e^{-N}\,.
\end{equation}
This is the hallmark that a {\it first-order phase transition} is taking place, which may ultimately lead to an exponentially large anneal time, preventing any possibility to solve the problem with quantum annealers. Indeed, exponentially closing gaps are a common bottleneck in adiabatic quantum computation \cite{10.1143/PTPS.184.290,Jorg_2010,SergeyZero2016} and a great deal of effort has been invested in finding ways to circumvent this problem, by adding, for example, additional drivers and/or catalysts, see Refs.~\cite{PhysRevE.85.051112, Seoane_2012, Nishimori2017,Albash2019,Takada_2021,Albash2021,Feinstein:2022xcc,Feinstein:2024cbd}. Additionally, the works mentioned provide an extensive description of the use of stoquastic catalysts in contrast with non-stoquastic ones, in particular whether or not the introduction of non-stoquastic interactions enhances the performance compared to the case of the traditional stoquastic formulation. We emphasise that in the rest of the paper we use the term ``stoquastic'' to refer to an operator with non-positive off-diagonal matrix elements in the computational basis; whereas the term ``non-stoquastic'' includes all the other cases.

In this paper we report a statistical analysis using multiple XX-catalysts on randomly generated MWIS problem instances comparing the performance with conventional quantum annealing. Using numerical approaches for diagonalising the Hamiltonian at each discretised anneal step, we find statistical evidence of a general gap enhancement when a stoquastic catalyst is employed, confirming the power of this class of catalysts for certain problems (see Refs.~\cite{Crosson2020designing,Nishimori2020} for similar results). This may open a route towards solving the MWIS problem in polynomial time using quantum annealers with multiple XX-catalysts, overcoming its complexity from the classical perspective.

The paper is organised as follows: after a brief introduction to the maximum weighted independent set problem in Sec.~\ref{sec:MWIS}, we move on to a comparative analysis in Sec.~\ref{sec:analysis} where we discuss the results and effectiveness of our proposed catalyst in randomly constructed MWIS problem instances on two different classes of networks. We finally conclude in Sec.~\ref{sec:final}.

\section{The MWIS problem on Quantum Annealers}
\label{sec:MWIS}

The Maximum Weighted Independent Set (MWIS) problem is a fundamental problem in graph theory and combinatorial optimisation which has a number of real-world applications, from network design \cite{zhao2012}, scheduling problems \cite{ARKIN1987} and combinatorial auctions \cite{DeVries2003}, to molecular biology \cite{BAFNA1996} and clustering aggregation \cite{Li2012}. In its simplest formulation, it consists in finding a subset of non-connected vertices $\widetilde{V}(G) \subset V(G)$ of a weighted graph $G$ such that the sum of the weights of the nodes in $\widetilde{V}(G)$ is maximised. It is well known that this problem is NP-hard, which means there is no known polynomial-time algorithm to solve it for general graphs. This makes it computationally challenging, especially for large graphs.

It can be shown that solving the MWIS problem on a weighted graph $G$ with $N$ vertices $V(G) = \{w_i \,|\, i=1,...,N \}$ --- where $w_i$ is the weight associated to the node $i$ --- and edges $E(G)$, is equivalent to finding the ground state of the Hamiltonian in Eq.~\eqref{eq:problem} with the following linear and quadratic couplings~\cite{choi2008minorembeddingadiabaticquantumcomputation,Choi:2010yxd}
\begin{align}
    h_i ~&=~ \sum_{j \in {\rm nbr}(j)} \, J_{ij} - 2 \, w_i \nonumber \\
    J_{ij} ~&\geq~  \min(w_i, w_j) \, ,
    \label{eq:MWIS_hJ}
\end{align}
where ${\rm nbr}(i) = \{j : (i,j) \in E(G)  \}$ is the subset of vertices directly connected with node $i$. The definition of the couplings in Eq.~\eqref{eq:MWIS_hJ} guarantees that the solution of the MWIS problem is encoded in the ground state of the Hamiltonian in Eq.~\eqref{eq:problem} and in our convention will correspond to the subset constituted by the spins pointing upwards.

\begin{figure}[t!]
\centering
\includegraphics[keepaspectratio, width=0.47
\textwidth]{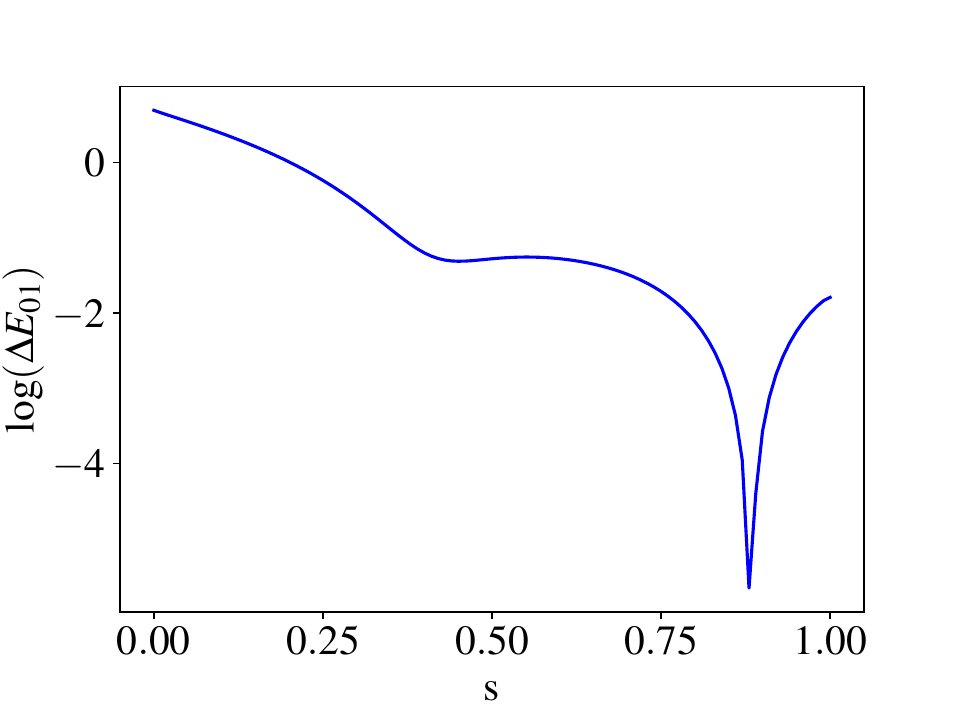}
\caption{The log of the energy difference between the ground and the first excited state versus the annealing parameter $s$ for a random generated MWIS problem instance with {\it 10} nodes. We can clearly see that it exhibits an exponentially closing gap around $s \approx {\it 0.9}$, indicating that a first-order phase transition is taking place.}
\label{fig:first_order}
\end{figure}

Being notoriously hard to solve classically, one may wonder if the MWIS problem may be efficiently tackled using quantum annealers, ultimately circumventing all the classical obstructions which prevent its solution in polynomial time. These obstructions are mainly due the exponential growth in the system size of the subsets of independent vertices which have to be checked and sorted by their total weight in order to solve the problem. One might expect significant advantages from using quantum annealing for this type of problem. Certain parameter regimes of the MWIS problem, however, exhibit first-order phase transitions during the adiabatic evolution of the system. As outlined in the introduction, this results in an exponential increase in anneal time required to ensure adiabatic evolution. In Fig.~\ref{fig:first_order} we report a typical example of a first-order phase transition in a randomly generated MWIS problem instance on a random graph where an exponentially closing gap between the ground state and the first excited state appears around $s \approx 0.9$.

Among the various approaches adopted to remove first-order phase transitions or to mitigate their effects, we shall focus on a particular one that involves the use of catalysts. Catalysts in quantum annealing can be designed to modify the energy landscape, enhancing the minimum energy gap and thereby facilitating the adiabatic evolution of the system. By carefully selecting and tuning these catalysts, it is possible to significantly mitigate the challenges posed by first-order phase transitions. In the following we shall investigate on how a particular class of catalysts leads to a general gap improvement on randomly generated MWIS problem instances.

\section{Adding a catalyst term in the Hamiltonian}
\label{sec:analysis}

Before going through the details of our analysis, let us provide a general overview on how catalysts can be used in quantum annealing. As mentioned in the previous section, a catalyst is usually introduced to avoid or mitigate the effect of first-order phase transitions. Formally, it corresponds to a set of new interactions appearing in the annealing Hamiltonian $H_A$ in Eq.~\eqref{eq:annealing_hamiltonian}, which now reads
\begin{equation}
    H_A ~=~ (1-s)\,H_D \,+\, s \, H_P \, + \, s(1-s) \, H_C \, ,
\end{equation}
where the catalyst term $H_C$ is in general a combination of the $X$,$Y$ and $Z$-fields. The catalyst term does not have any effect at the beginning and at the end of the evolution, \ie, at $s=0$ and $s=1$, due to the prefactor $s(1-s)$. Instead, it affects the energy levels during the adiabatic evolution of the system with an impact that increases and decreases quadratically in the anneal parameter $s$.

In certain MWIS problem instances with a bi-partite structure one can observe that a single XX-catalyst defined as the product of two X-fields acting on two nodes of the graph is effective in enhancing the minimum energy gap \cite{Feinstein:2022xcc}. In addition, in Ref.~\cite{Ghosh:2024lqz} we show on how specific catalysts can completely remove first-order phase transitions in the same class of problems with specific characteristics. However, these approaches would at least require partial information on the specific instance, for example the partition structure, symmetries, general properties of the graph, {\it etc}. In contrast, here we will adopt a complementary approach by randomly generating thousands of problem instances and operating without any {\it a-priori} knowledge of the structural properties of the graphs. This method allows us to comprehensively assess the average efficacy of the proposed catalyst on the MWIS problem in its full generality. 

From now on we shall focus on a particular catalyst which will be at the centre of our entire analysis. We define
\begin{equation}
    H_C ~=~ J_c \, \sum_{i,j \in E(G)} \, \sigma_x^{(i)} \, \sigma_x^{(j)} \, ,
    \label{eq:XXcata}
\end{equation}
where we note that the summation is limited to $i$'s and $j$'s which are directly connected by an edge in the problem graph and where $J_c$ is a parameter that tunes the strength of the catalyst. We note that this catalyst has already been adopted in Ref.~\cite{Albash2019,Takada_2021} though in different contexts from the one we shall analyse. We will we set $J_c = -1$ for the rest of this paper, meaning that we are introducing a {\it stoquastic} catalyst. Many works have been directed to the study of non-stoquastic catalysts, see for example Refs.~\cite{Seki2012,Seoane_2012_2,Crosson:2014gud,Seki_2015,Nishimori2017_2, PhysRevB.95.184416, Albash2019}, demonstrating their efficacy in improving the minimum energy gap in various scenarios. However, parallel studies have proved that stoquastic catalysts can lead to similar improvements under certain conditions for specific problem settings \cite{Crosson2020designing, Nishimori2020, Ghosh:2024lqz}. We will see later in this section that our results are in agreement with these latter findings, demonstrating that a stoquastic catalyst represents the preferred choice with respect to its non-stoquastic counterpart in the case of MWIS problems on random graphs.

In particular, the catalyst in Eq.~\eqref{eq:XXcata} consists of a set of XX-fields ``attached'' at each pair of vertices which defines the graph structure of the MWIS problem under study. In the following we shall focus on two classes of graphs: random graphs and scale-free networks --- picking Erdős–Rényi graphs and Barab\a'asi-Albert networks as archetypal examples of the two categories, respectively. In both cases we will generate MWIS problem instances on randomly generated graphs with random parameters uniformly distributed in the following ranges
\begin{align}
    J_{ij} ~\equiv~ J ~&\in~ [1,2] \nonumber \\
    w_i ~&\in~ [0,1] \, ,
    \label{eq:ranges}
\end{align}
where we choose the couplings $J_{ij}$ to be constants, \ie, independent of the sites $i$ and $j$, such that the second line of Eq.~\eqref{eq:MWIS_hJ} is always satisfied. Note that this can be done without any loss of generality, \ie, without restricting the number of possible realisations of the MWIS problem.

Let us now describe in more detail the kind of networks which will serve as a testing ground to benchmark the effectiveness of the catalyst in Eq.~\eqref{eq:XXcata}.

\vspace{0.5cm}
\subsection{Erdős–Rényi Graphs}

Erdős-Rényi (ER) graphs are a fundamental model in the study of random graphs, widely applied across fields such as computer science, physics, biology, and social sciences. An Erdős-Rényi graph $G(N,p)$ is defined by two parameters: the number of vertices $N$ and the probability $p$ with which each pair of vertices is connected by an edge. The parameter $p$ in this model can be thought of as a weighting function; as $p$ increases the model becomes more and more likely to include graphs with more edges.

\begin{figure*}
\centering
\subfloat[][$p ~=~ 0.25$]{
\includegraphics[width=0.5\textwidth]{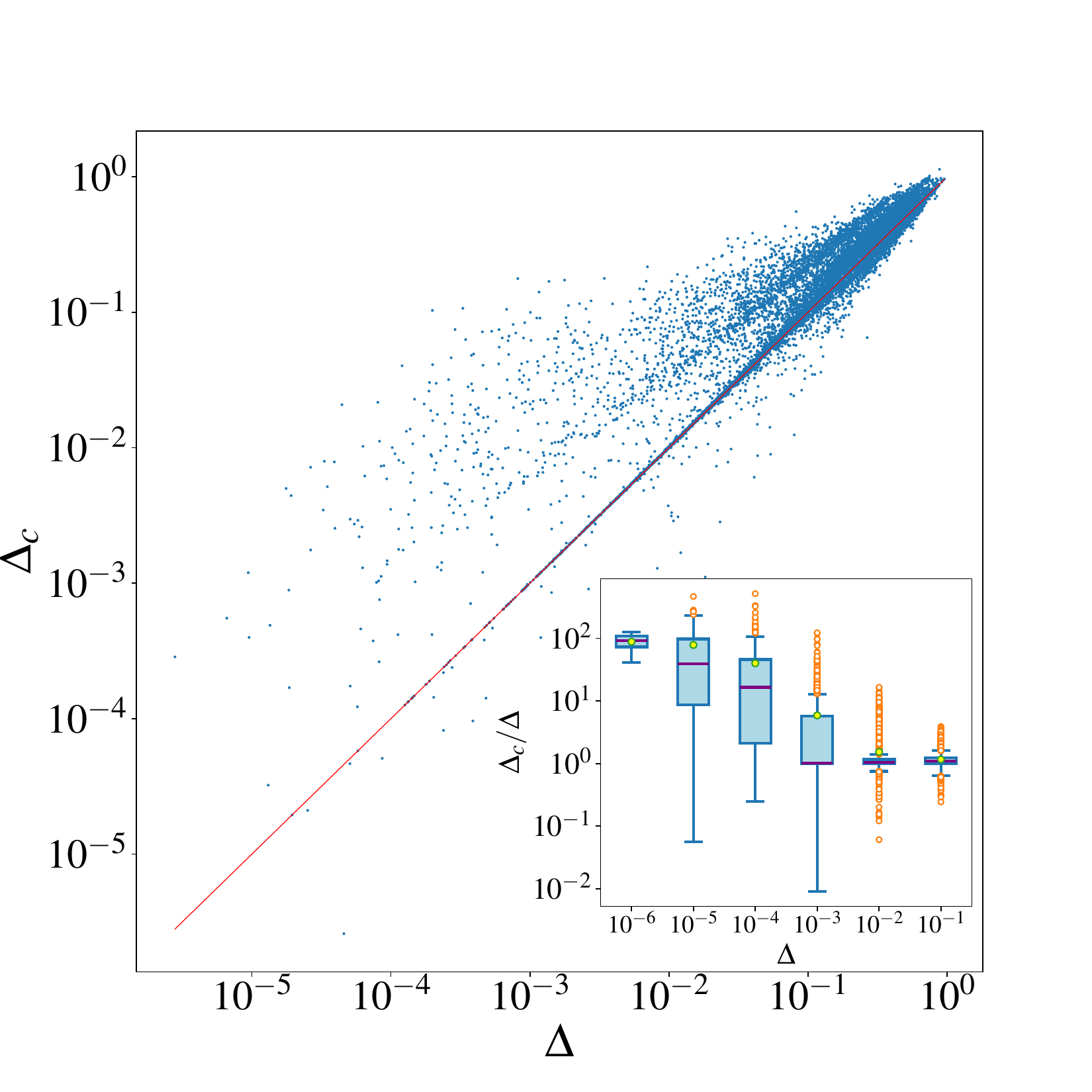}}
\subfloat[][$p ~=~ 0.4$]{
\includegraphics[width=0.5\textwidth]{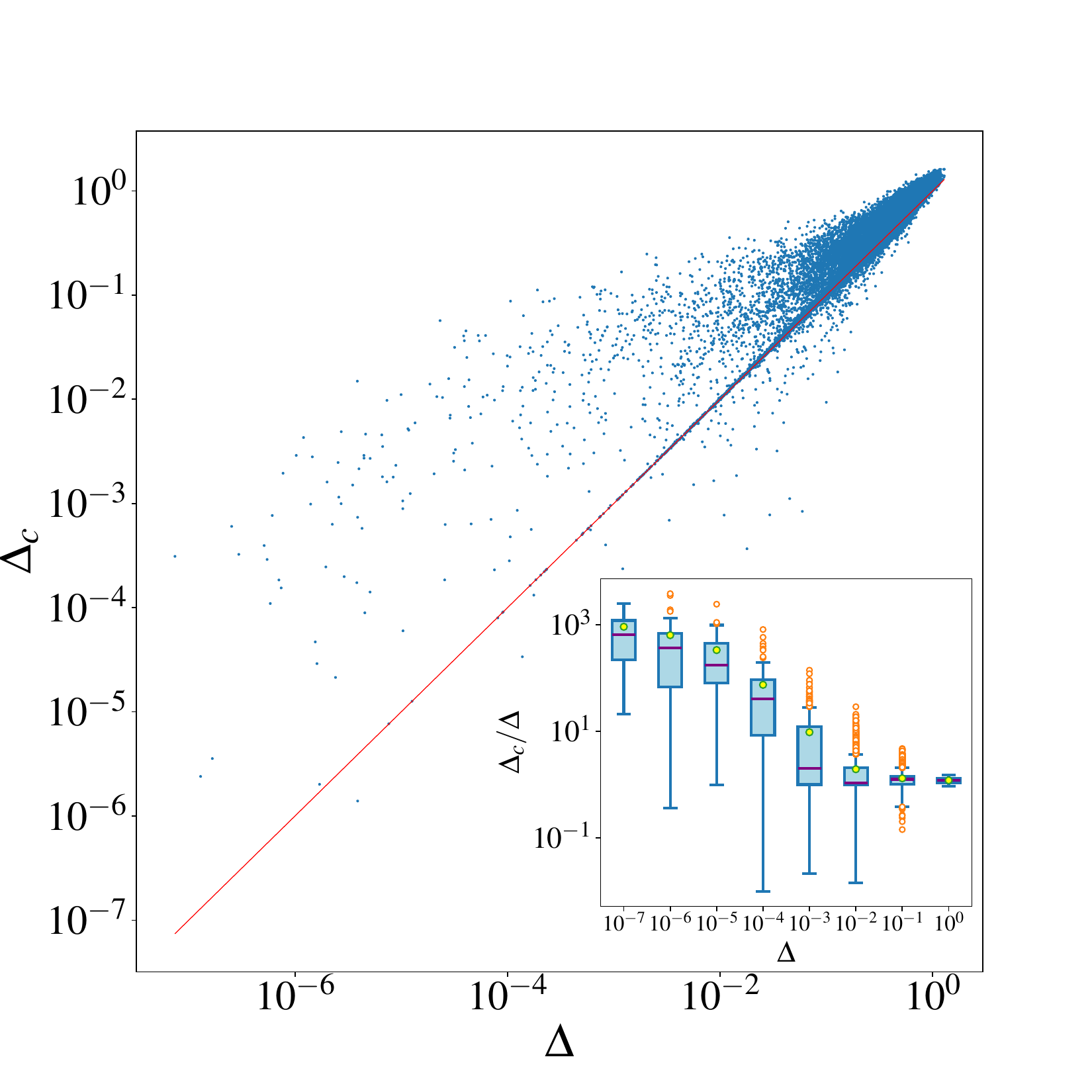}}\\
\subfloat[][$p ~=~ 0.6$]{
\includegraphics[width=0.5\textwidth]{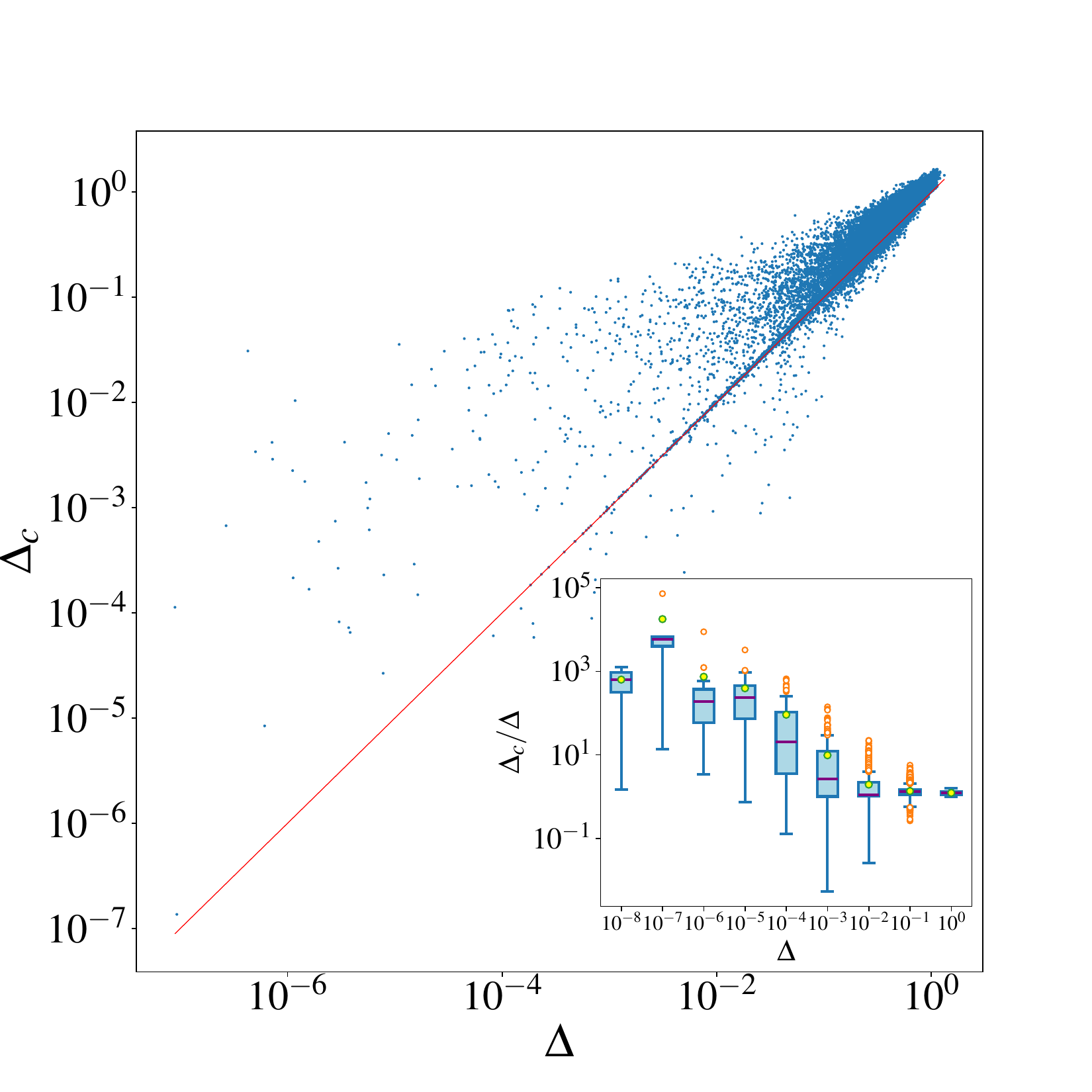}}
\subfloat[][$p ~=~ 0.8$]{
\includegraphics[width=0.5\textwidth]{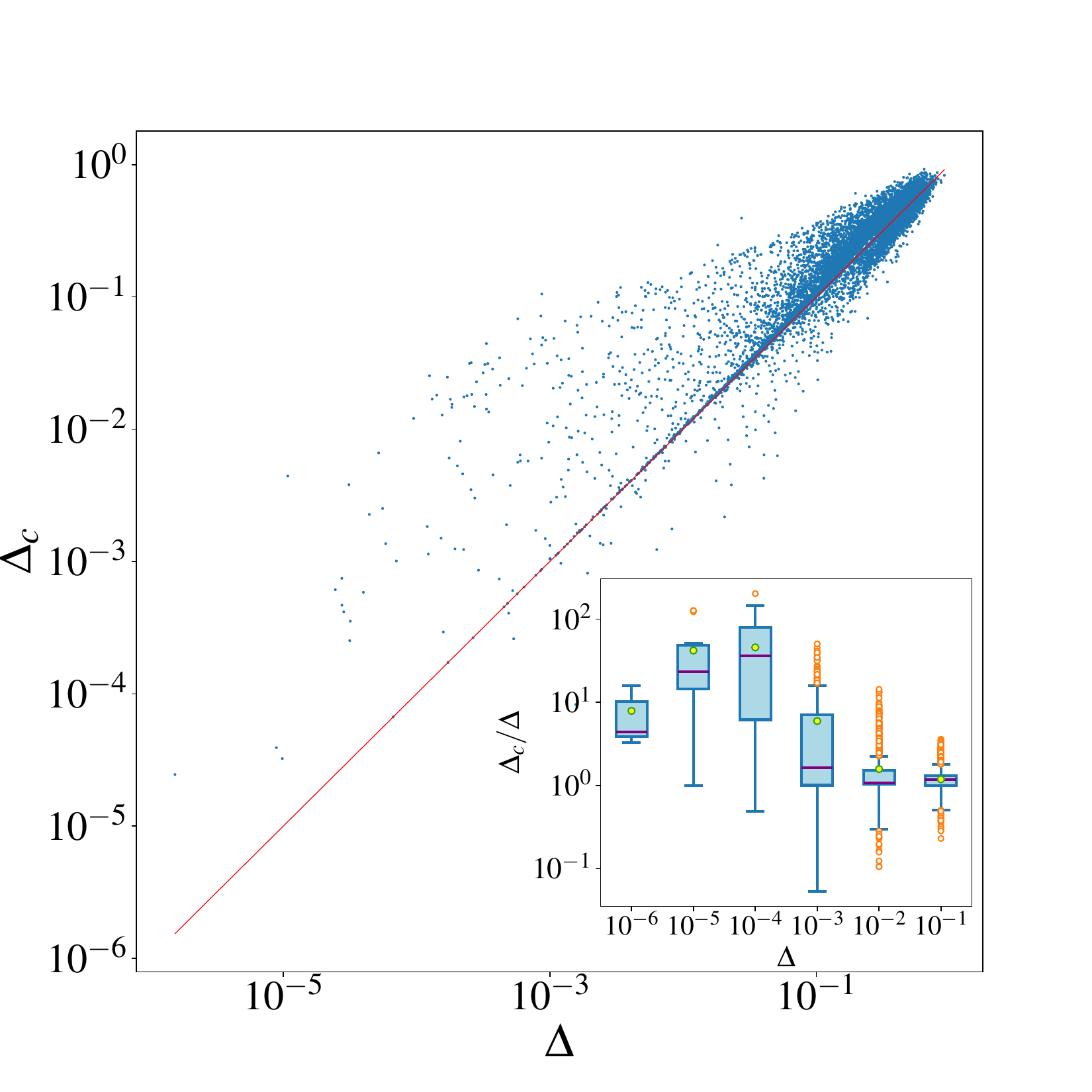}}
\caption{Randomly generated MWIS problem instances  on Erdős–Rényi networks for different values of edge probability $p$. For each of these values we generated more than 10,000 instances and plotted $\Delta_c$, the minimum gap with multiple XX-catalyst versus the minimum gap, $\Delta$, in the catalyst-free case. In all these cases we observe a general gap enhancement, meaning that the majority (${\it 74\%}$ on average) of the instances lay above the critical red line $\Delta_c = \Delta$ which divides the region in which the catalyst enhances the minimum gap from the region in which its effect is opposite. The box plots show the distribution of the ratio $\Delta_c / \Delta$ for various orders of magnitude of the original minimum energy gap $\Delta$. The purple line and the yellow circle indicate the median and the mean respectively. The top and bottom edges of the box show the upper and lower inter-quartile range. Outliers, depicted as orange circles, are determined if they are more than 1.5 times more than the inter-quartile range away from the median. The two caps attached to each box show the maximum and minimum data points, excluding outliers. The labels ${\it 10^{-n}}$ on the horizontal axis must be interpreted as a shorter notation to indicate the range from ${\it 10^{-n}}$ to ${\it 10^{-(n+1)}}$.}
\label{fig:erdosrenyi}
\end{figure*}

To conduct a statistical analysis, we have generated random Erdős–Rényi graphs with $N=10$ nodes and constructed random MWIS problems on them, using parameters uniformly chosen from the ranges specified in Eq.~\eqref{eq:ranges}. In particular we have analysed more than 10,000 problem instances each for four different values of $p$ and compared the minimum energy gap with and without the catalyst. In Fig.~\ref{fig:erdosrenyi} we can see the effect of using this catalyst for $p = 0.25, 0.4, 0.6, 0.8$, showing that this method enhances the minimum energy gap for most of the instances. Of particular interest is the central region of these plots, where the approach without the catalyst leads to a minimum energy gap between $10^{-5}$ and $10^{-3}$, indicating the potential presence of a first-order phase transition. The instances in this region receive a gap enhancement when the multiple XX-catalyst is used, increasing the minimum gap by up to three orders of magnitude, eliminating the first-order phase transition in several cases. To better assess the effectiveness of this catalyst, we have added an inset to each figure showing a box plot representing  the distribution of the gap improvement, measured as $\Delta_c / \Delta$, for various orders of magnitude of the minimum energy gap $\Delta$. The purple line and the yellow circle represent the median and the mean, respectively. From the box plots we note that harder instances, populating the left hand region of each scatter plot, are distributed around larger values of $\Delta_c / \Delta$ meaning that these instances get a larger gap improvement with respect to easier problems with bigger minimum gaps, confirming what one can qualitatively deduce from the corresponding scatter plot.

We also stress that the same results cannot be achieved using the non-stoquastic version of the same catalyst, \ie, setting $J_c > 0$ in Eq.~\eqref{eq:XXcata}. Although the scenario that emerges using the non-stoquastic version of the same catalyst is comparable to the one depicted in Fig.~\ref{fig:erdosrenyi} for harder instances (meaning problems with $\Delta < 10^{-3}$), we also observe that for $\Delta$ approaching values of $10^{-2}$, the non-stoquastic catalyst leads, on average, to a further shrinkage of the minimum gap, becoming disadvantageous in this regime. Indeed, most of the instances lay below the critical line $\Delta_c = \Delta$, meaning $\Delta_c < \Delta$ in the majority of the cases. We have analysed the non-stoquastic case for different positive values of $J_c$, all leading to similar results of the case discussed in Appendix~\ref{sec:app1}, where we show and discuss the results for $J_c = 1$.

\subsection{Barab\a'asi-Albert graphs}

Barab\a'asi-Albert (BA) graphs constitute a different class of graphs than the previous scenario as they are ``scale-free''. This arises from the fact that the distribution of degrees across different vertices (nodes) of this graph follows a power law decay, distinct from, for example, the normal distribution decay in the previous case. This in turn implies that several of the nodes in such a graph have a preferential attachment to many other nodes, and are called hubs. This introduces a new kind of complexity in the problem compared to the Erdős–Rényi networks.

Apart from the theoretical motivation, studying MWIS problems in these graphs might have a strong impact on real-world problems. All large scale networks such as citation networks, the internet or social networks have a BA graph structure~\cite{redner1998popular,Albert-Diameter-1999, Baraba_si_1999}. In principle, we can weigh each node by its degree, quantifying the ``importance'' of an individual in the network and then the MWIS problem will help identify the most important individuals in the network who have no connection with each other. This would be greatly useful to model and understand social interactions. 

A Barabási-Albert network is defined through a parameter $m$, which determines the number of edges that each new node introduces when it is added to the network. Essentially, $m$ represents how many connections a new node will form with existing nodes. When a new node is added to the network, it does not connect to existing nodes randomly. Instead, it follows the principle of preferential attachment, meaning that nodes with higher degrees (more connections) are more likely to receive new links. This mechanism ensures that well-connected nodes continue to attract more connections, leading to the formation of hubs. The choice of $m$ significantly impacts the structure of the network. A larger $m$ results in a network with higher average connectivity, as each new node introduces more connections. 

\begin{figure*}[]
\centering
\subfloat[][$m ~=~ 3$.]{
\includegraphics[width=0.5\textwidth]{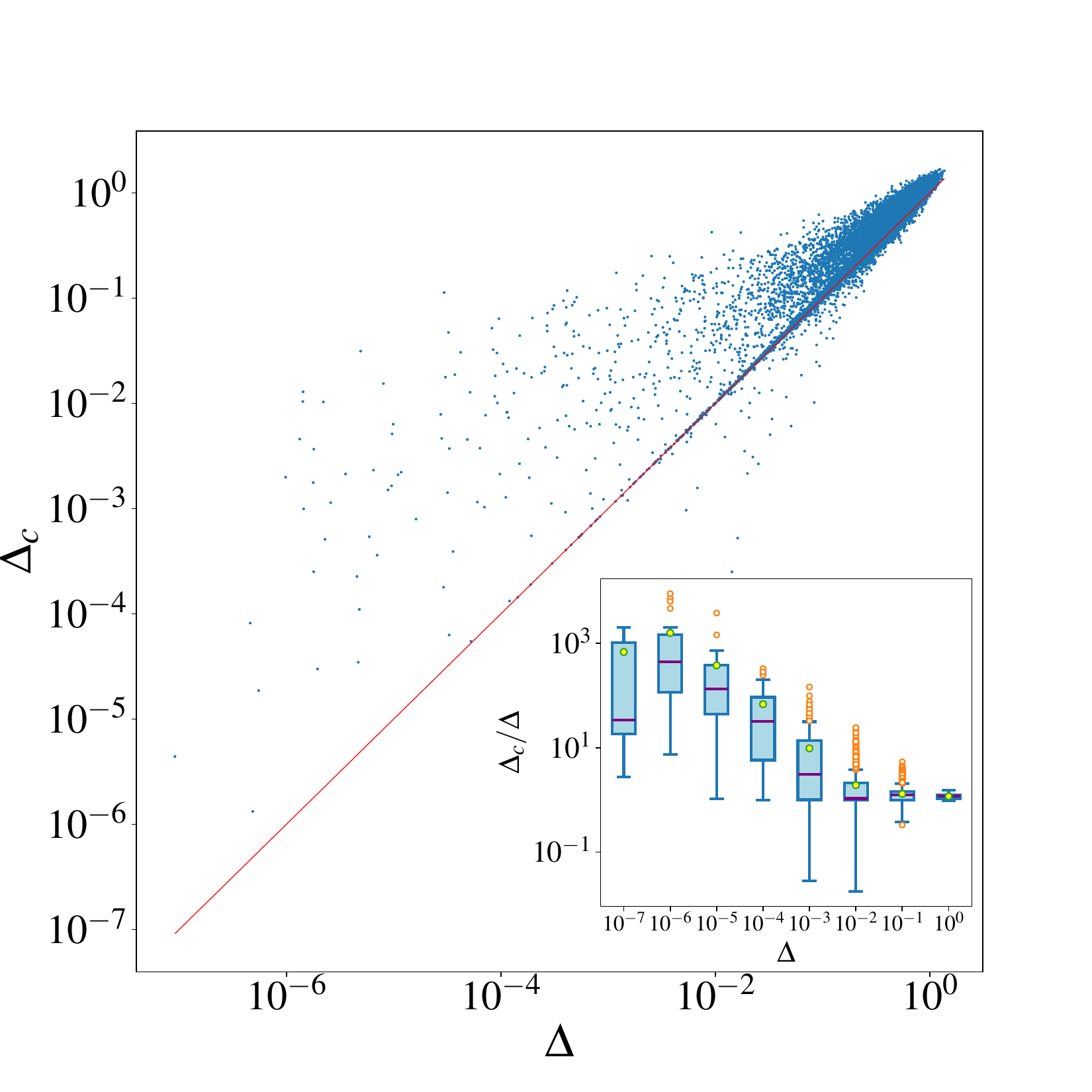}}
\subfloat[][$m ~=~ 4$.]{
\includegraphics[width=0.5\textwidth]{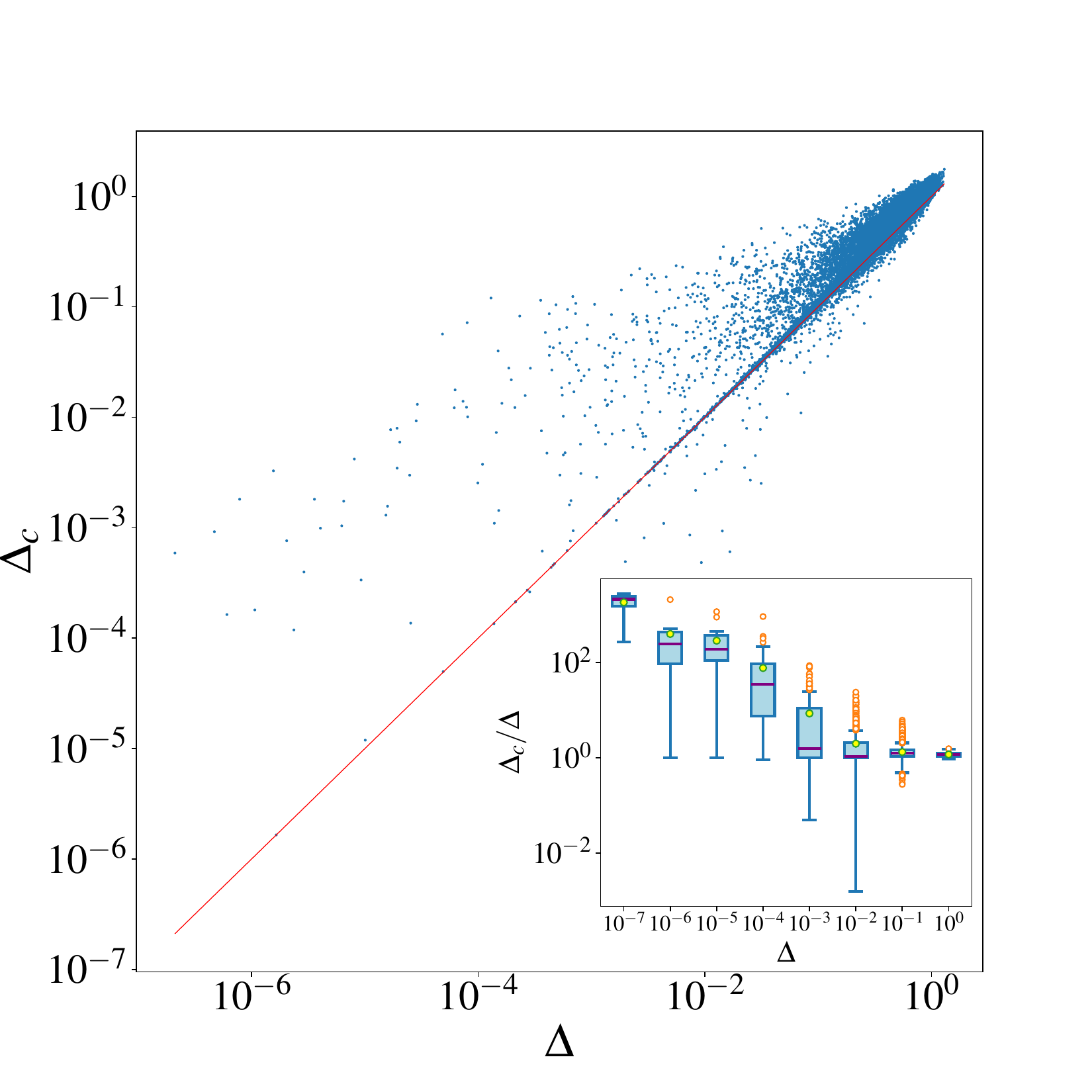}}\\
\subfloat[][$m ~=~ 5$.]{
\includegraphics[width=0.5\textwidth]{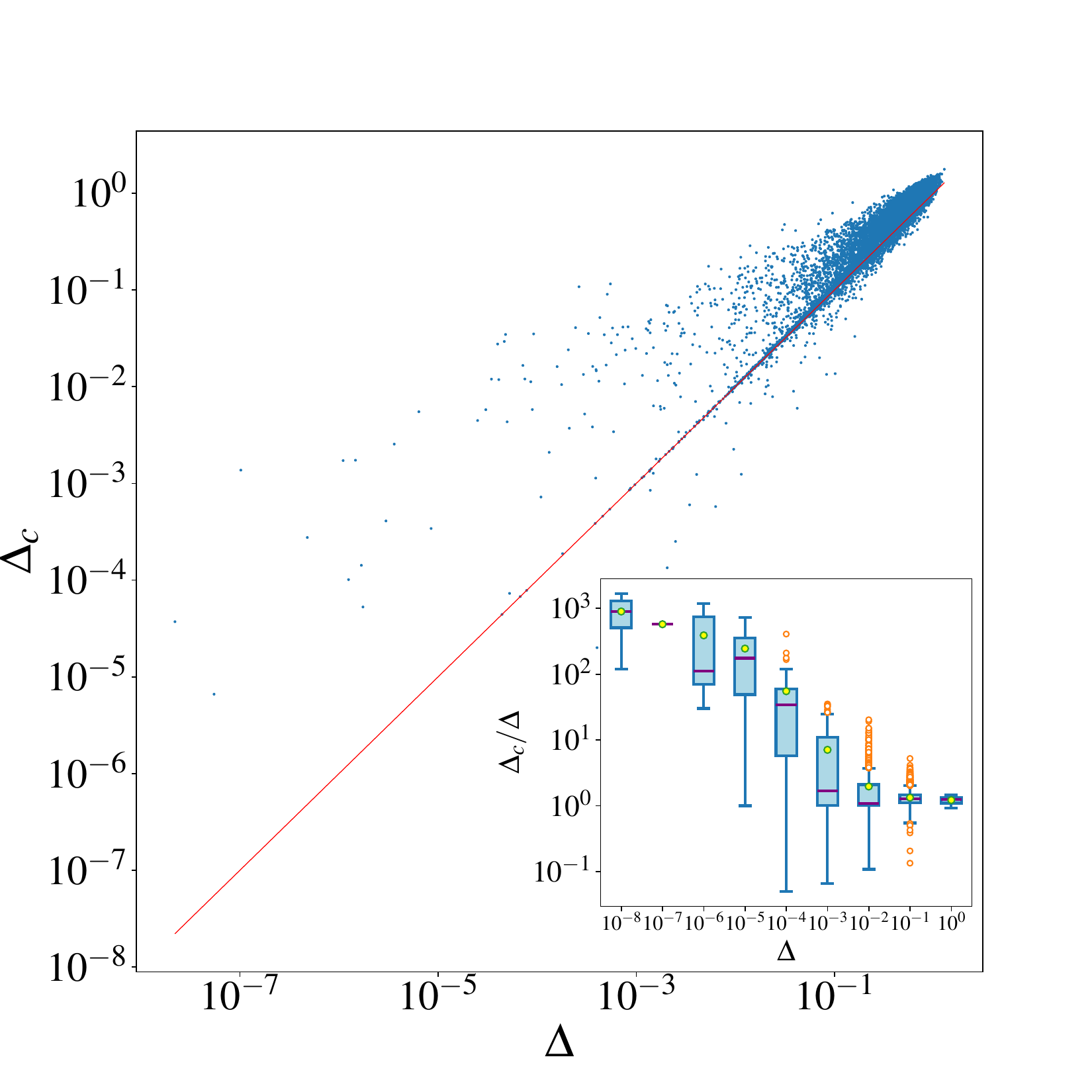}}
\subfloat[][$m ~=~ 6$.]{
\includegraphics[width=0.5\textwidth]{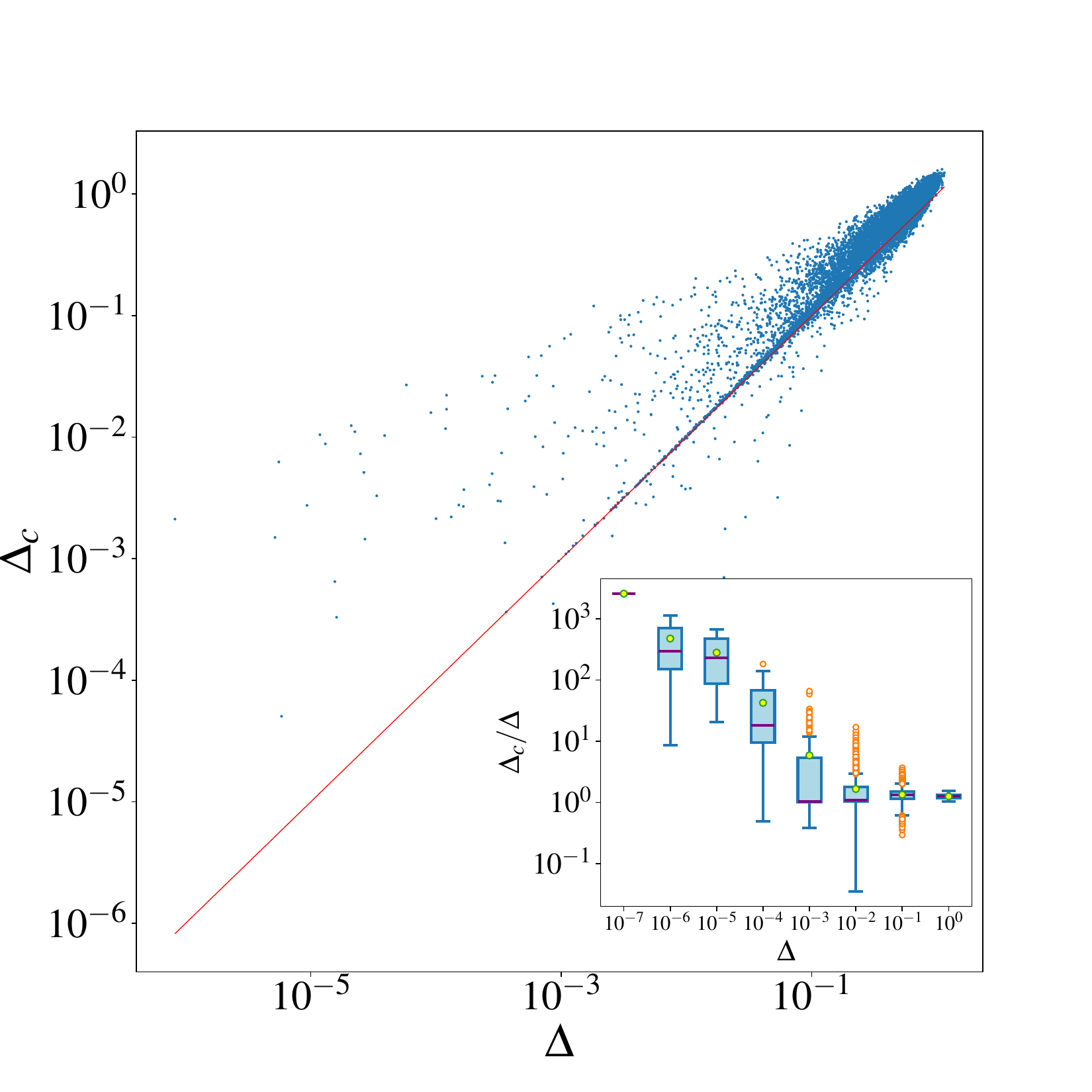}}
\caption{Randomly generated MWIS problem instances on Barab\a'asi-Albert networks for different values of $m$. For each of these values we generated more than 10,000 instances and plotted $\Delta_c$, the minimum gap with multiple XX-catalyst versus the minimum gap, $\Delta$, in the catalyst-free case. In all these cases we observe a general gap enhancement, meaning that the majority of the instances (${\it 81\%}$ on average) lay above the critical red line $\Delta_c = \Delta$ which divides the region in which the catalyst enhances the minimum gap from the region in which its effect is opposite. The box plots show the distribution of the ratio $\Delta_c / \Delta$ for various orders of magnitude of the original minimum energy gap $\Delta$. The purple line and the yellow circle indicate the median and the mean respectively. The top and bottom edges of the box show the upper and lower inter-quartile range. Outliers, depicted as orange circles, are determined if they are more than 1.5 times more than the inter-quartile range away from the median. The two caps attached to each box show the maximum and minimum data points, excluding outliers. The labels ${\it 10^{-n}}$ on the horizontal axis must be interpreted as a shorter notation to indicate the range from ${\it 10^{-n}}$ to ${\it 10^{-(n+1)}}$.}
\label{fig:barabasialbert}
\end{figure*}

In what follows, we take the same approach as before by putting random weights and random couplings as specified in Eq.~\eqref{eq:MWIS_hJ}. We then asses the efficiency of the proposed catalyst on this new set of graphs in the same way we did for the case of Erdős–Rényi networks, \ie, generating more than ten thousand problem instances for four different values of $m$ and comparing the minimum energy gap with and without the catalyst. In Fig.~\ref{fig:barabasialbert} we show the result for $m = 3,4,5,6$. We note that this catalyst is again statistically effective in enhancing the minimum energy gap. The region where it is reasonable to think that the evolution goes through a first-order phase transition --- say $\Delta \lesssim 10^{-2}$ --- is populated by points which are mostly distributed above the diagonal, meaning that the first order transition is less dramatic or is even removed by the catalyst. As before, the insets show the distribution of the gap improvement for different orders of magnitude of $\Delta$, leading to the same conclusion as in the Erdős–Rényi case. We again stress that these results cannot be achieved using the non-stoquastic version of this catalyst, \ie, setting $J_c > 0$ in Eq.~\eqref{eq:XXcata}. Indeed, the scenario which emerges is the same as discussed in the Erdős–Rényi case. See Appendix~\ref{sec:app1} for a more detailed discussion.

\subsection{Scaling properties}

\begin{figure}[]
\centering
\includegraphics[keepaspectratio, width=0.47
\textwidth]{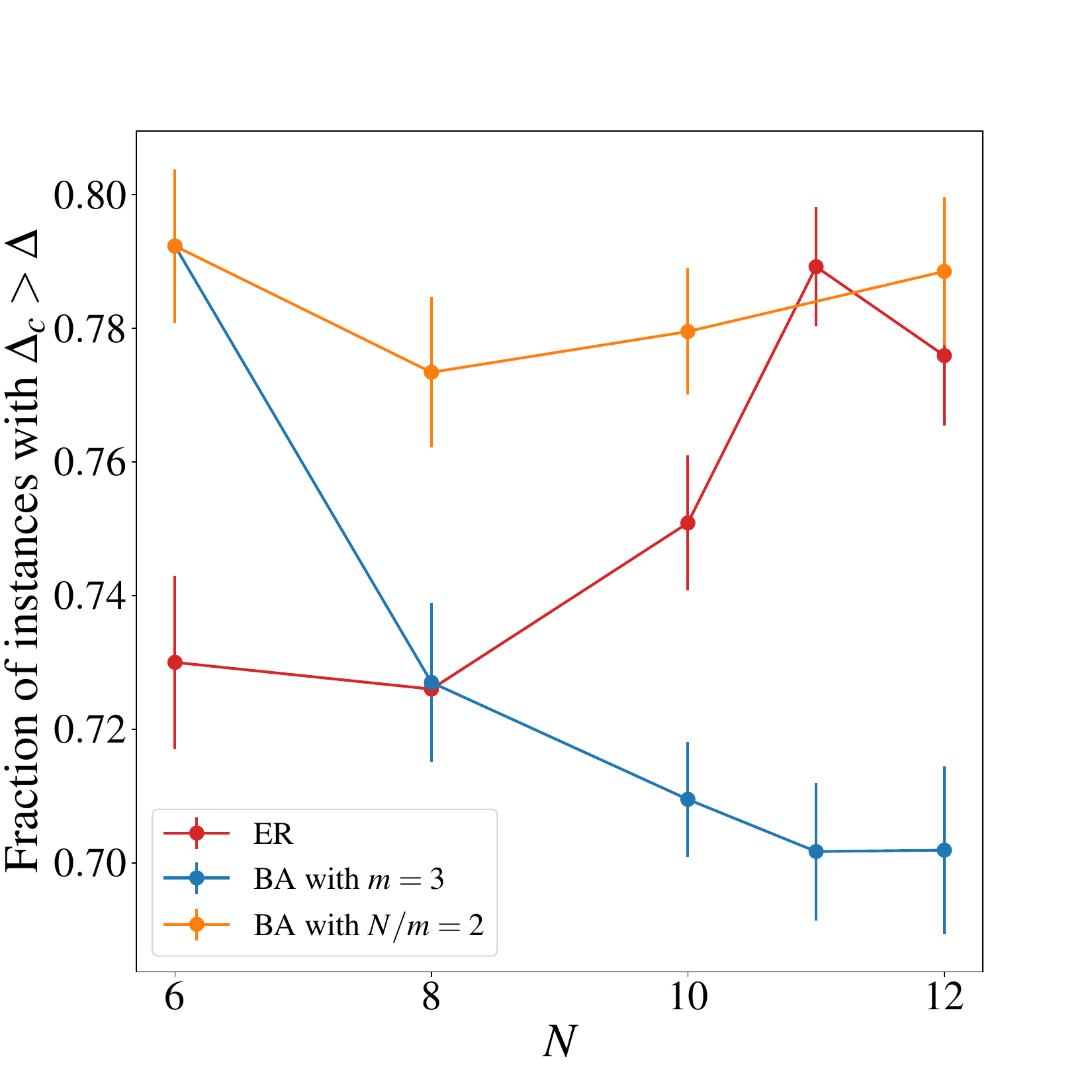}
\caption{Fraction of potential hard (\ie, with $\Delta < 10^{-1}$) MWIS problem instances on BA and ER graphs which benefit from the use of multiple XX-catalysts versus $N$, the total number of nodes. We have performed the analysis for both Erdős–Rényi with $p=0.5$ (red) and Barab\a'asi-Albert with $m=3$ constant (blue) and $m$ variable such that $N/m = 2$ (orange). For each point in the graph we generated more than ten thousand problem instances. The associated error bars are computed via bootstrapping methods, resampling from the original dataset.}
\label{fig:scaling}
\end{figure}

\begin{figure}[]
\centering
\includegraphics[keepaspectratio, width=0.47
\textwidth]{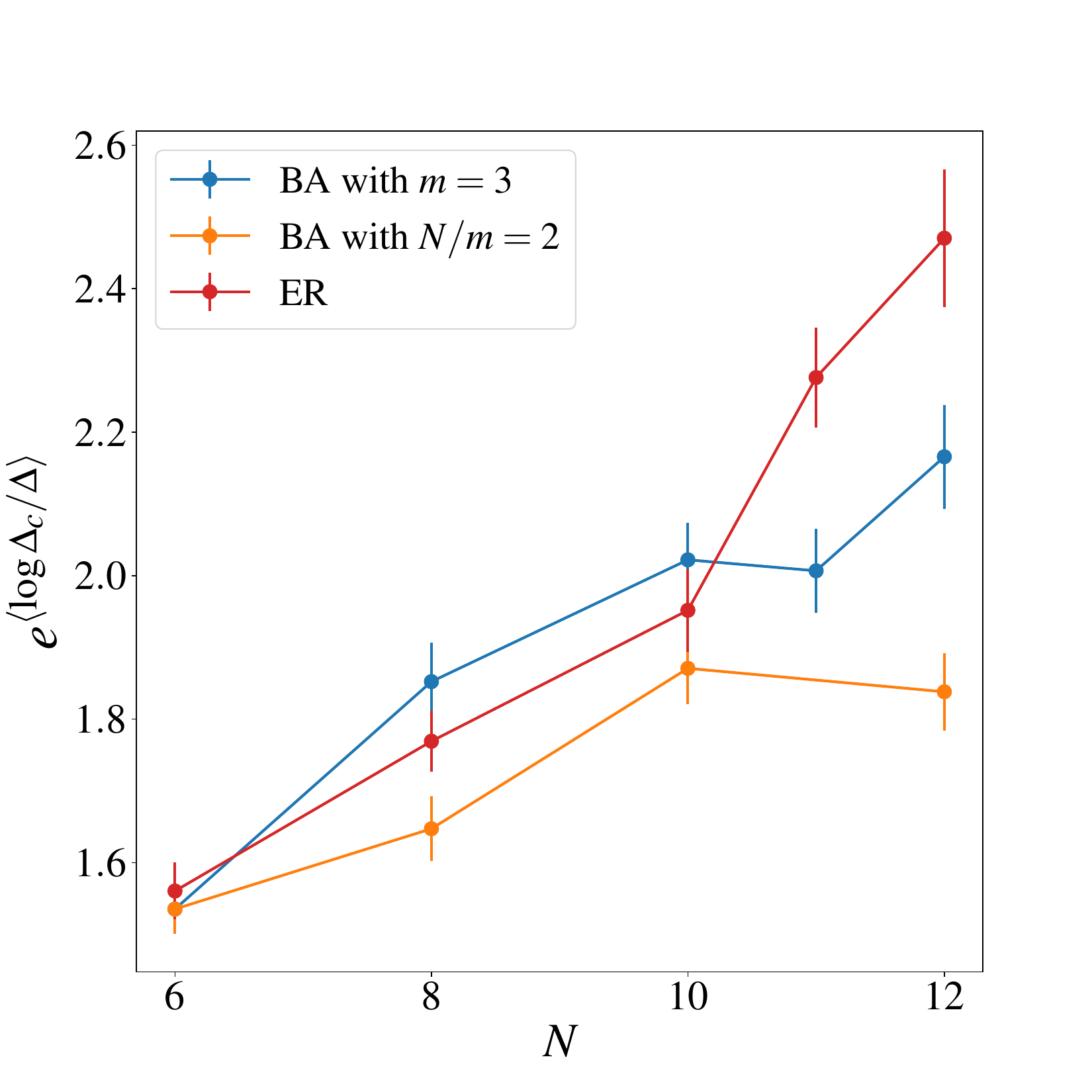}
\caption{Average catalyst gap improvement computed as the geometric mean of $\Delta_c / \Delta$ versus total number of vertices in the graph. In computing this average we used over ten thousand MWIS problem realisations for each data point, taking into account instances with $\Delta < 10^{-1}$. The error bars are computed taking into account how the error on $\langle \Delta_c / \Delta \rangle$ propagates.}
\label{fig:scaling2}
\end{figure}

Having shown statistical evidence that a general gap enhancement can be achieved using multiple XX-catalysts in randomly generated MWIS problems with $N=10$ nodes, we now consider how the average improvement scales with the total number of nodes in the graph. In Fig.~\ref{fig:scaling} we have plotted the fraction of potential harder instances (meaning those ones with $\Delta < 10^{-1}$) which receive a gap improvement from the catalyst (\ie, the fraction of instances with $\Delta_c > \Delta$) for various numbers of nodes, where $\Delta_c$ and $\Delta$ are the minimum gaps with and without the catalyst, respectively. Additionally, in Fig.~\ref{fig:scaling2} we show the same scaling plot but with a different metric, denoted as the average gap improvement and quantified as the geometric mean of the ratio $\Delta_c / \Delta$. Since this ratio represents an improvement rate, the geometric mean is a more meaningful estimator than the arithmetic mean. For the Erdős–Rényi case the data are collected by generating random graphs with $p=0.5$, whereas for the Barab\a'asi-Albert we analysed two cases: one with $m=3$ independently of the number of the spins $N$ and one maintaining a constant ratio $N/m = 2$. Fig.~\ref{fig:scaling} shows that the majority of the randomly generated MWIS problem instances present a larger minimum energy gap once the catalyst is used. We note a monotonic decreasing behaviour only in the BA case with constant $m$, while the other two cases present oscillations of the order of a few percent. Given the limited number of data points it is not possible to extrapolate the behaviour for large $N$ in the three cases analysed, which certainly needs to be addressed in a future study.

On the other hand, Fig.~\ref{fig:scaling2} displays an overall increasing behaviour in terms of the average gap improvement for both scale-free and random networks as the number of total nodes in the graph increases. Even more notably, the scaling persists although the fraction of harder instances ---\ie, those ones with $\Delta_c < 10^{-3}$, increases with $N$ as well, tripling from $N=6$ to $N=12$  in both random and scale-free networks. 
Finally, we now give a brief explanation about why, in general, this catalyst is efficient in enhancing the gap between the ground and the first excited state of the problem Hamiltonian in most of the MWIS problem instances analysed. In Ref.~\cite{Ghosh:2024lqz} we discuss how an informed choice of the catalyst may have a huge impact on the minimum energy gap, with the possibility of removing a first-order phase transition in several cases. In particular, it turns out the efficacy of a catalyst strongly depends on how its composition reflects the structure of the ground and first excited states of the problem Hamiltonian in terms of their Hamming distance. Indeed, it turns out that to maximise the effect of the catalyst (which would correspond to a complete removal of the first-order phase transition), one would need to introduce a catalyst which includes all the spin flip operators that connect the ground and the first excited state of $H_P$. This of course would require knowledge of the structure of the ground state which is as hard as solving the optimisation problem itself. However, one can achieve satisfactory results without any a-priori knowledge of the structure of the global minimum or eventual symmetries of the problems. This is indeed the scenario which is emerging from the analysis conducted in this paper, where the catalyst in Eq.~\eqref{eq:XXcata} is able to improve the minimum energy gap in the majority of the MWIS problem realisations, regardless of all the features that the specific instance has. In this context, this means that the catalyst used throughout this whole analysis introduces (on average) a sufficient number of spin flip operators to connect, at least perturbatively, the ground and first excited states in a statistically significant number of MWIS problem cases. 

\section{Discussion and conclusions}
\label{sec:final}

In this paper, we have conducted a study of the enhancement of the minimum energy gap in Maximum Weighted Independent Set (MWIS) problems using multiple XX-catalysts. Our findings provide robust statistical evidence that the introduction of this catalyst can lead to significant gap improvements in randomly generated MWIS problem instances. This improvement is crucial as it potentially mitigates the bottleneck of exponentially closing gaps, which are a common problem in adiabatic quantum computation.

For our analysis, we randomly generated thousands of MWIS problem instances on two types of graphs: Erdős–Rényi and Barab\a'asi-Albert, both of which have widespread real-world applications. Firstly, we fixed the total number of nodes in the graphs at $N=10$. From this analysis, we found that almost 80\% of the instances exhibited a noticeable improvement in the gap due to the introduction of the catalyst, regardless of the type of graph --- whether scale-free or random --- on which the MWIS instance was constructed. Interestingly, this analysis revealed that the catalyst is more effective on specific problem instances that feature smaller minimum gaps, $\Delta$. Specifically, the catalyst demonstrates greater efficacy on instances with potential first-order phase transitions, characterised by gaps $\Delta \lesssim 10^{-2}$. For these harder instances, the average improvement ratio, $\Delta_c / \Delta$, shows a significant increase, suggesting that the catalyst specifically enhances performance on these more challenging cases. Conversely, instances with larger minimum gaps do not experience the same degree of improvement. This selective targeting of more complex instances highlights the potential of this catalyst as a powerful tool for addressing hard-to-solve cases, where the presence of a first-order phase transition plays a critical role in determining performance outcomes.

Secondly, we investigated on how the efficacy of this catalyst scales with the problem size, from $N=6$ to $N=12$ nodes. The results show that the catalyst consistently improves the minimum energy gap, leading to $\Delta_c > \Delta$ from $70\%$ to almost $80\%$ of the cases, depending on the class of the MWIS graph. Due to limited computational capabilities, it has not been possible to investigate the regions beyond $N=12$. Furthermore, it is not even feasible to reliably extrapolate the scaling behaviours for larger $N$ because of the presence of substantial statistical fluctuations. These fluctuations can introduce large oscillations in the data, obscuring any clear trends or patterns, and making it difficult to draw meaningful conclusions about the underlying scaling behavior based solely on the available information.

By contrast, analysing the average gap improvement $\exp \langle \log \Delta_c / \Delta \rangle$, we observe a notable trend: this quantity consistently increases as the problem size grows. What is particularly striking is that this behaviour holds even as the fraction of more challenging problems --- those likely to exhibit a first-order phase transition --- rises significantly with increasing $N$. Specifically, the proportion of these harder instances nearly triples, going from $N=6$ to $N=12$, yet the average gap improvement continues to expand. This suggests a robust scaling property that persists despite the increasing difficulty associated with larger problem sizes.

Overall, these results highlight the significant potential of stoquastic catalysts in enhancing quantum annealing performance for certain combinatorial optimisation problems, giving to stoquastic catalysts a privileged role with respect to their non-stoquastic counterpart in this specific scenario. Indeed, our analysis on MWIS problems reveals that contrary to the stoquastic approach, the non-stoquastic version of this catalyst becomes statistically counterproductive as soon as the minimum energy gap $\Delta$ approaches values of the order of $10^{-2}$, which could still correspond to a reservoir of potentially hard problem instances.

While our current computational capabilities limited the exploration to graphs with up to $N=12$ nodes, the trends observed in the case of a stoquastic catalyst suggest similar improvement can be achieved for larger graphs.
This development could potentially open up a new path for solving MWIS problem instances in polynomial time using quantum annealers.

\vspace{1.5cm}
\begin{acknowledgments}
LAN, RG, SB and PAW are supported by EPSRC grant EP/Y004590/1 ``MACON-QC''.
\end{acknowledgments}

\appendix 
\section{Results using a non-stoquastic multiple XX-catalyst}
\label{sec:app1}

\begin{figure*}
\centering
\subfloat[][$p ~=~ 0.4$.]{
\includegraphics[width=0.5\textwidth]{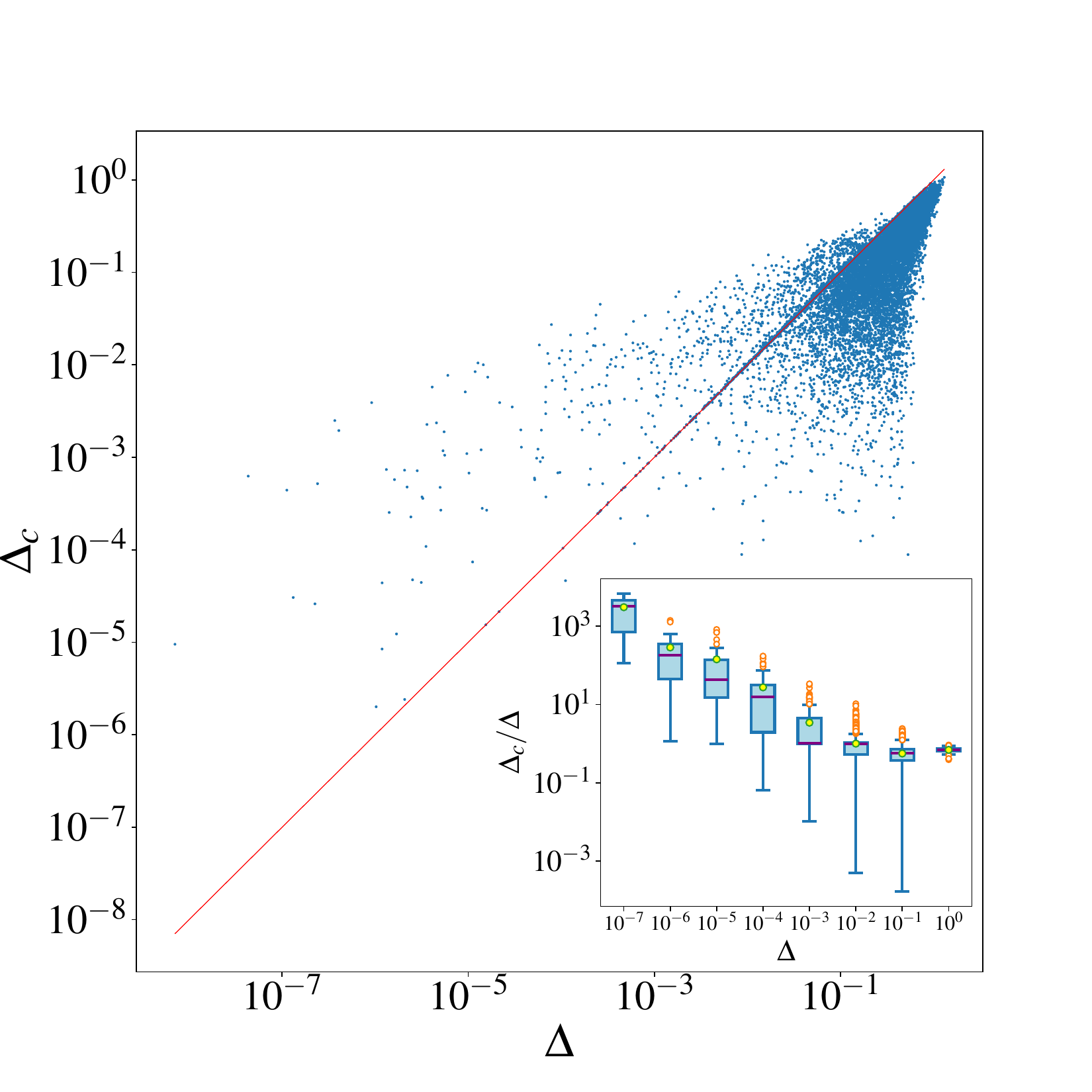}}
\subfloat[][$p ~=~ 0.6$.]{
\includegraphics[width=0.5\textwidth]{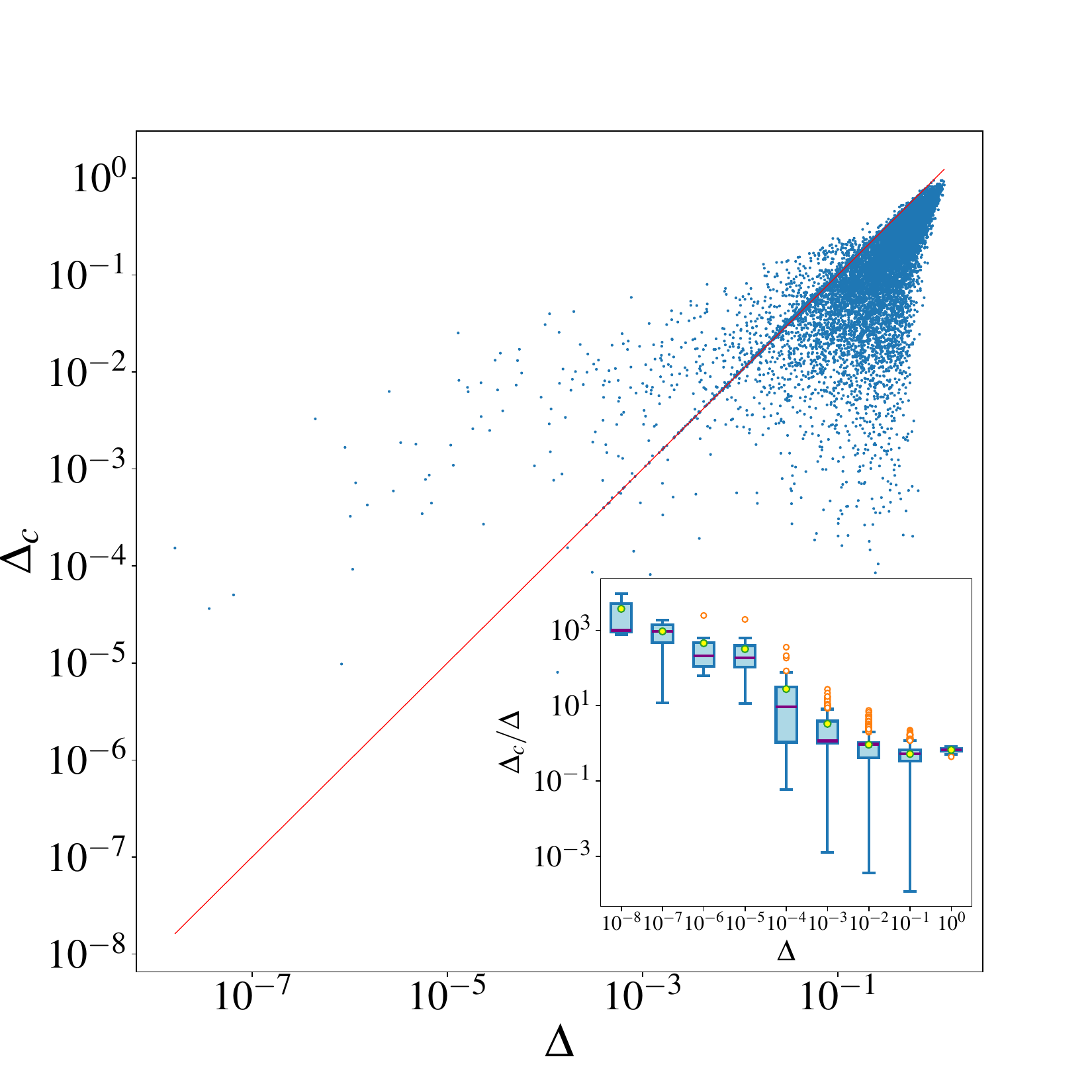}}\\
\subfloat[][$m ~=~ 4$.]{
\includegraphics[width=0.5\textwidth]{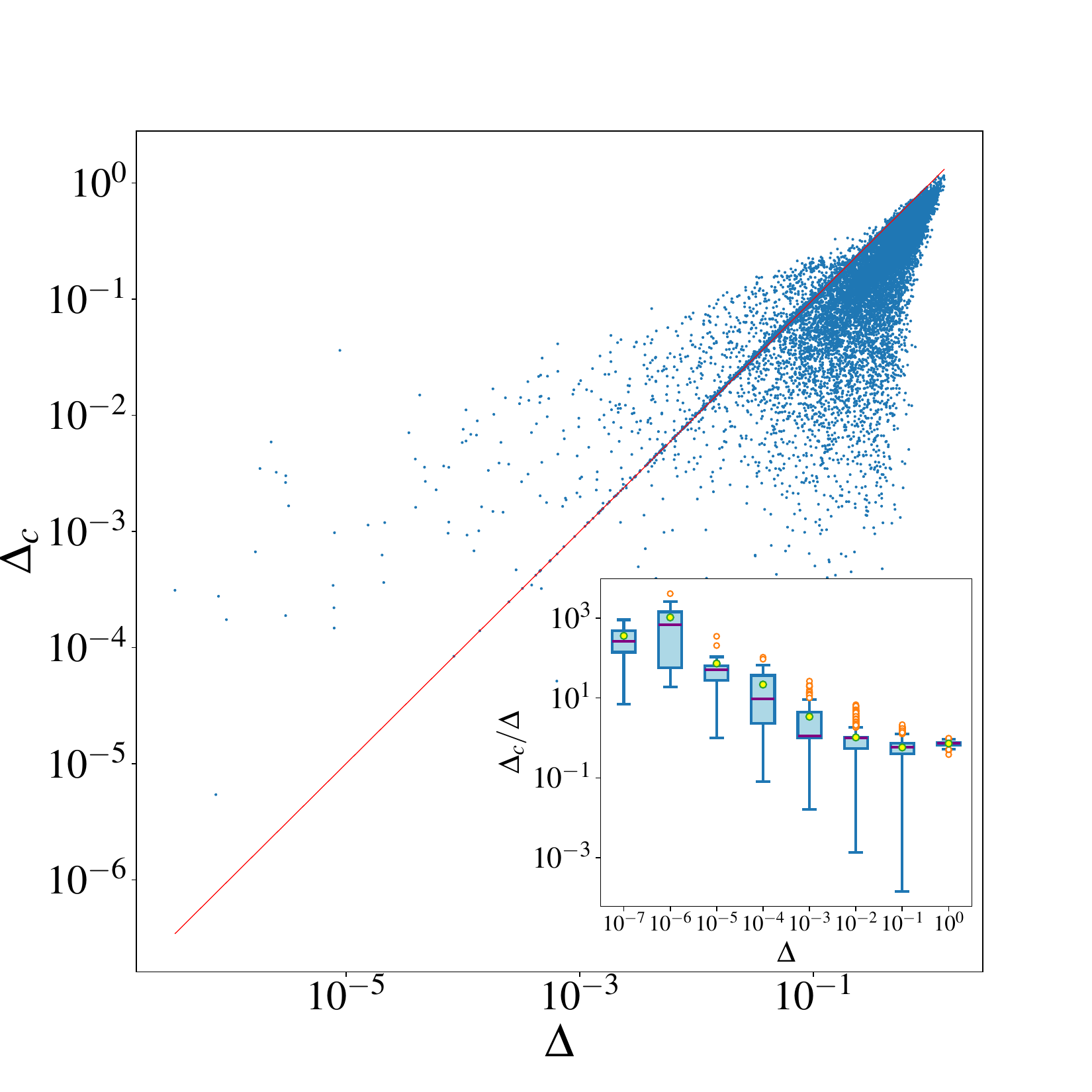}}
\subfloat[][$m ~=~ 5$.]{
\includegraphics[width=0.5\textwidth]{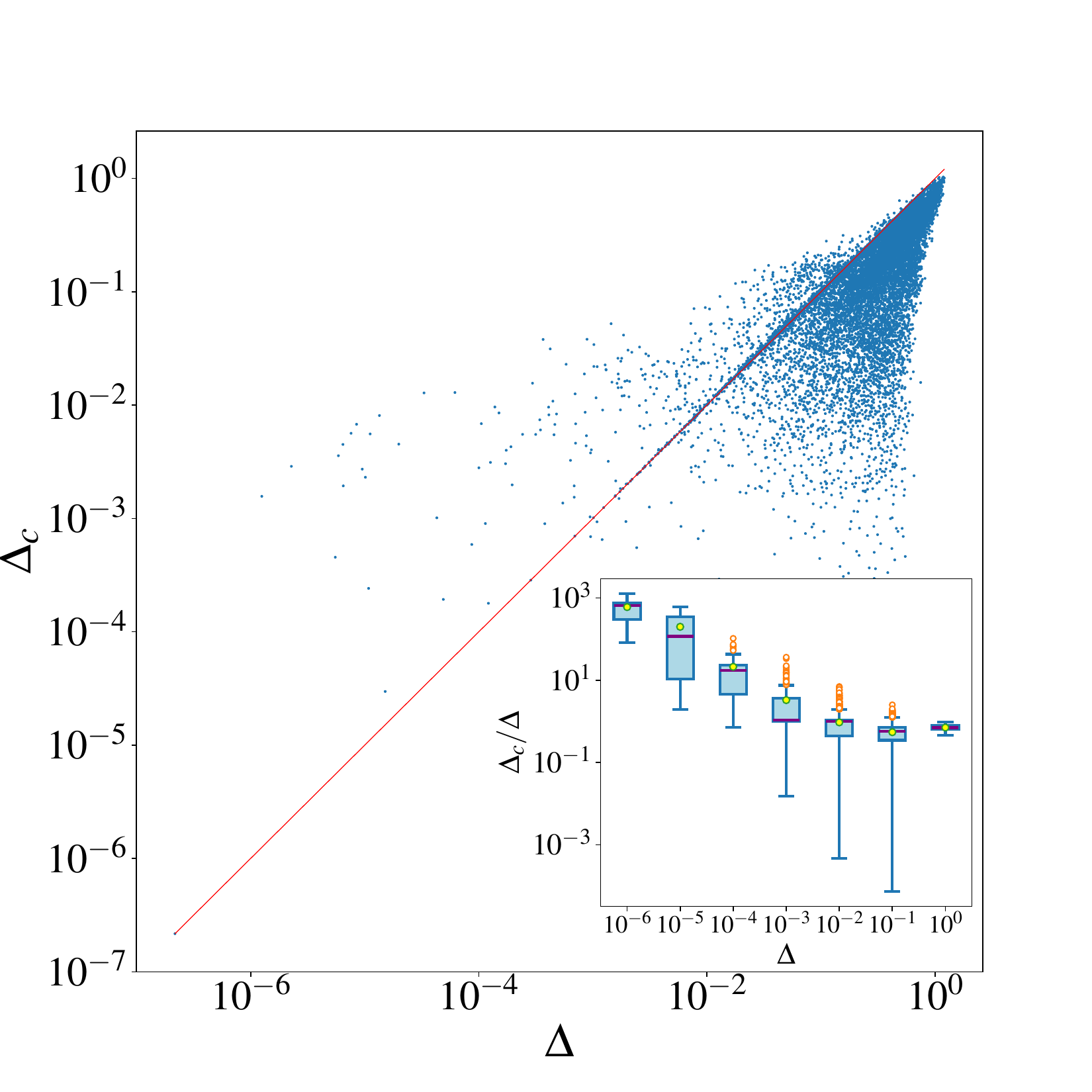}}
\caption {Comparison of the minimum energy gap using a non-stoquastic version of the catalyst (setting $J_c = 1$) with the minimum energy gap in the catalyst free case on randomly generated MWIS problem instances constructed on both Erdős–Rényi, in (a) and (b), and Barab\a'asi-Albert graphs, in (c) and (d). While harder instances (those with $\Delta < 10^{-3}$) still benefit from a gap improvement comparable to what is achieved with the stoquastic version, it is important to highlight a different trend for instances with moderately larger minimum gaps. These instances, particularly those with gaps $\Delta \gtrsim 10^{-2}$, experience a reduction in the gap due to the catalyst, as is evident from the insets. This suggests that although the catalyst is still effective for more difficult problems, it can negatively impact the performance on relatively easier instances by shrinking their gaps. As a consequence, more than 90\% of the MWIS realisations lie in the region below the red line, contrary to what happens in the stoquastic case. }
\label{fig:NS}
\end{figure*}

In Fig.~\ref{fig:NS}, we present the analogous results to Figs.~\ref{fig:erdosrenyi} and \ref{fig:barabasialbert}, but for the non-stoquastic version of the catalyst defined in Eq.~\eqref{eq:XXcata}. For this analysis, we set $J_c = 1$ and selected $p = 0.4, 0.6$ for Erdős–Rényi graphs and $m = 4, 5$ for Barabási–Albert networks, both with $N=10$, as representative examples. In this scenario, we observed that approximately 90\% of the data points fall below the red line, $\Delta_c = \Delta$, indicating that the majority of instances experience a reduction in the minimum energy gap after applying the catalyst. This suggests that, unlike the stoquastic version, the non-stoquastic catalyst tends to reduce the minimum gap for most instances.

Interestingly, choosing other positive values of $J_c$ did not result in substantial changes compared to the case shown here. Despite the overall gap shrinkage, we note that harder problem instances, specifically those with $\Delta \lesssim 10^{-3}$, still exhibit an improvement in their minimum gap, comparable with the behaviour seen in the stoquastic version. On the other hand, a notable feature in all cases is the presence of a tail of instances originating from the top-right corner of each figure and extending into the region where $\Delta_c < \Delta$. This is further confirmed by the box plots, which display a pronounced tail corresponding to data points in the ranges $10^{-2} \div 10^{-1}$ and $10^{-1} \div 10^{0}$. These instances show significant reductions in the gap ratio, with $\Delta_c / \Delta$ values dropping as low as $10^{-3}$.

The pronounced downward tail, especially in instances with larger gaps, highlights the limitations of the non-stoquastic catalyst compared to its stoquastic counterpart. Although the non-stoquastic catalyst provides comparable improvements for harder instances with $\Delta < 10^{-3}$, it becomes disadvantageous when the minimum gap approaches the order of $10^{-2}$. In these cases, it leads to a significant reduction in the gap ratio, $\Delta_c / \Delta$, which can even increase the likelihood of triggering a first-order phase transition. This makes the non-stoquastic catalyst less suitable for broad applicability, as it negatively impacts instances with moderately larger minimum gaps.

The data suggest that while the non-stoquastic catalyst retains some effectiveness for the most challenging instances, its detrimental effect on instances with slightly larger gaps limits its utility. In contrast, the stoquastic version appears more versatile, delivering consistent improvements across a wider range of problem complexities without the risk of such significant performance degradation for moderately hard problems. 

\bibliography{TheLiterature}

\end{document}